\documentclass[twocolumn,showpacs,preprintnumbers,amsmath,amssymb,prb,floatfix]{revtex4}

\usepackage{graphicx}

\begin{document}

\title{A low-noise ferrite magnetic shield}

\author{T.~W.~Kornack}
\email{tkornack@princeton.edu}
\author{S.~J.~Smullin}
\author{S.-K.~Lee}
\author{M.~V.~Romalis}
\affiliation{Physics Department, Princeton University, Princeton, NJ 08544, USA} 
\date{\today}

\begin{abstract}
Ferrite materials provide magnetic shielding performance similar to commonly used high permeability metals but have lower intrinsic magnetic noise generated by thermal Johnson currents due to their high electrical resistivity. Measurements inside a ferrite shield with a spin-exchange relaxation-free (SERF) atomic magnetometer reveal a noise level of 0.75~fT~Hz$^{-1/2}$, 25 times lower than would be expected in a comparable $\mu$-metal shield. We identify a $1/f$ component of the magnetic noise due to magnetization fluctuations and derive general relationships for the Johnson current noise and magnetization noise in cylindrical ferromagnetic shields in terms of their conductivity and complex magnetic permeability.
\end{abstract}

\maketitle

Many sensitive magnetic measurements depend on high performance magnetic shields typically made from high magnetic permeability metals.\cite{Mager:1970} Several layers of such $\mu$-metal can attenuate external fields by many orders of magnitude. The most sensitive measurements, however, are limited at the level of 1--10~fT~Hz$^{-1/2}$ by the magnetic noise generated by the innermost layer of the magnetic shield itself.\cite{Nenonen:1996,Kominis:2003} Superconducting magnetic shields do not generate magnetic noise,\cite{Vanthull:1967}  but thermal radiation shields required for their use with room-temperature samples typically also generate noise of 1--3~fT~Hz$^{-1/2}$.

MnZn ferrites are promising materials for magnetic shielding because of their high relative permeability ($\mu \sim 10^4 \mu_0$) and much higher electrical resistivity ($\rho \sim 1 \ \Omega$~m) than $\mu$-metal. We measured magnetic fields inside a 10~cm diameter MnZn ferrite shield using a spin-exchange relaxation free (SERF) atomic magnetometer and found that the magnetic noise level is up to 10 times lower than the noise measured in Ref.~3\nocite{Kominis:2003} from a 40~cm diameter $\mu$-metal shield. In addition to electrical resistivity other sources of dissipation can lead to magnetic noise in accordance with fluctuation-dissipation theorem. In particular, magnetic viscosity effects result in an imaginary component of magnetic permeability at low frequency and generate magnetization noise with a $1/f$ power spectrum.\cite{Vitale:1989} The measured low frequency magnetic noise inside the ferrite is in good agreement with a prediction for this magnetization noise based on independently measured complex permeability of the ferrite material. To aid with the design of low noise magnetic shields we also derive simple analytic relationships for Johnson current noise and magnetization noise inside infinitely long cylindrical magnetic shields.

\begin{figure}
\includegraphics{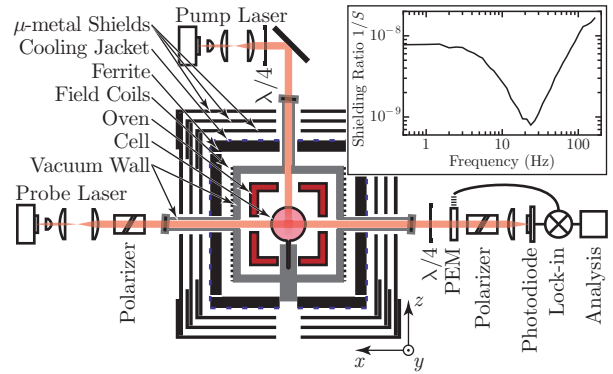}
\caption{(Color online) Schematic of the experimental apparatus. Three layers of $\mu$-metal and one layer of ferrite form a set of cylindrical shields, symmetric about the axis of the pump laser. The shielding factor of the shields was measured using a SERF magnetometer, formed in the K cell at the intersection of the pump and probe beams.}
\label{fig:schem}
\end{figure}

Shield performance was measured using a SERF magnetometer, diagrammed in Fig.~\ref{fig:schem}, that is similar to the design in Ref.~3\nocite{Kominis:2003}. A $\sim 2$~cm diameter spherical glass cell contains potassium in natural abundance, 3~atm of $^4$He to reduce wall relaxation, and $\sim 50$~torr N$_2$ gas to quench the K excited state. The cell was heated by a resistive twisted-pair wire heater driven by AC current and the K density was measured to be $3\times 10^{13} \ \mathrm{cm}^{-3}$, corresponding to a temperature of 170~C. The space around the oven is evacuated to $\sim 10$~millitorr for thermal insulation and laser light travels through the shields in evacuated tubes to reduce optical noise due to convection. The magnetometer is calibrated and tuned using a set of orthogonal field coils wrapped around a fiberglass vacuum enclosure sitting inside the ferrite shield. Cooling water keeps the ferrite well below its Curie temperature of $\sim 150$~C. Circularly polarized light tuned to the $D_1$ resonance of potassium at 770~nm polarizes the potassium atoms. Atomic spin precession in a magnetic field is detected by optical rotation of off-resonant, linearly polarized light. The polarization angle is precisely measured using a photoelastic modulator (PEM) and a lock-in amplifier.\cite{Zavattini:2006} The optical components are mounted on a breadboard less than 100~cm in diameter; the compact construction is intended to decrease low-frequency noise due to thermal drift.

Cylindrical shields are composed of three layers of 1.6~mm thick, high permeability $\mu$-metal surrounding a 10~mm thick, 104~mm inner-diameter ferrite shield. The shields have 18~mm access holes along the three principal axes of the device. The ferrite shield is assembled from three parts: an annulus and two discs as endcaps. After degaussing the assembled shields, the residual field is typically found to be 0.1--2~nT and is compensated by the internal field coils. The quasi-static magnetic shielding factor $S$ of the ferrite layer was measured to be 170 in the radial direction and 100 in the axial direction. Whereas the shielding factor in the radial direction agrees with calculations based on the $\mu$ of the ferrite, the shielding factor in the axial direction is somewhat lower than expected, which is likely due to small gaps between the ferrite parts. The mating ferrite surfaces had been polished to $\sim 1$~$\mu$m flatness to allow for tight sealing of the shielding enclosure, thereby reducing loss in the shield's magnetic circuit. The quasi-static shielding factor of the assembled ferrite and $\mu$-metal shields was measured by the magnetometer to be $1.2 \times 10^{8}$ in the $y$ direction, as shown in Fig.~\ref{fig:schem}. An enhancement of the shielding factor at 22~Hz is due to the cancellation of the external field leaking through the holes in the shields with the phase-shifted field penetrating through the shield.\cite{Mager:1970}

\begin{figure}
\includegraphics{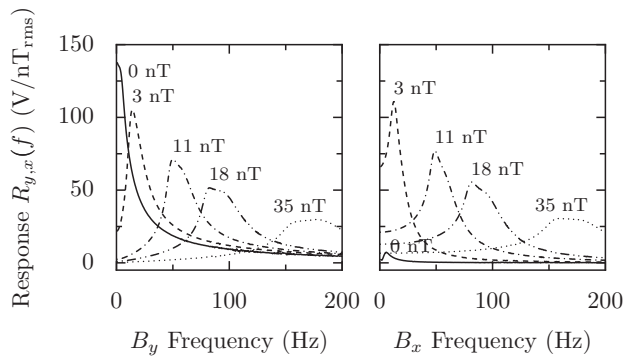}
\caption{Calibration response functions $R_y(f)$ and $R_x(f)$ of the magnetometer to excitations of $B_y$ and $B_x$ with static $B_z$ fields applied as indicated next to each trace. Response to $B_z$ excitation is strongly suppressed. For clarity, data for only five of the recorded $B_z$ values are shown. At higher values of $B_z$, broadening of the resonance is due to varying gyromagnetic ratio from inhomogenous polarization.\cite{Savukov:2005} Note that the applied $B_z$ field labels are approximate.} 
\label{fig:response}
\end{figure}

Measurements of the ferrite shield magnetic noise performed with the SERF magnetometer were limited by optical rotation noise outside the magnetometer bandwidth of $\sim 10$~Hz. In order to measure the magnetic noise spectrum over a larger frequency range, we tuned the magnetometer to resonate at higher frequencies by applying a constant $B_z$ field. The resonant frequency of the atomic spins is given by $f = \gamma B_z/2 \pi q$ where $\gamma=g \mu_B / \hbar$ is the gyromagnetic ratio for the electron and $q$ is the slowing-down factor due to the hyperfine interaction and spin-exchange collisions in the SERF regime.\cite{Savukov:2005} The magnetometer noise spectrum was calibrated by applying known fields in 3 directions and measuring the magnetometer response as a function of frequency. The response curves $R_y(f)$ and $R_x(f)$ to modulated $B_y$ and $B_x$ are shown in Fig.~\ref{fig:response}; the response to modulated $B_z$ is negligible, $R_z(f) \simeq 0$. As the figure indicates, in the limit of low magnetic field $B$, a SERF magnetometer is sensitive only to the field $B_y$ perpendicular to the plane of the lasers. When the field parallel to the pump beam $B_z$ corresponds to a Larmor frequency greater than the atomic relaxation rate, the magnetometer also has a significant resonant response to the field $B_x$ parallel to the probe beam; the response in this regime is comparable in magnitude to the response to $B_y$.\cite{Li:2006} Thus, the measured noise spectrum $V_m(f)$ is composed of the magnetometer response to noise in both $x$ and $y$ directions and can be calibrated using $\delta B_m(f)=V_m(f)/\sqrt{R_x(f)^2+R_y(f)^2}$. Note that we assume the noise in both $x$ and $y$ directions is the same due to the symmetry of the shield geometry and $\delta B_m(f)$ is the magnetic field noise along a single axis and is directly comparable to the theoretical calculations below.

\begin{figure}
\includegraphics{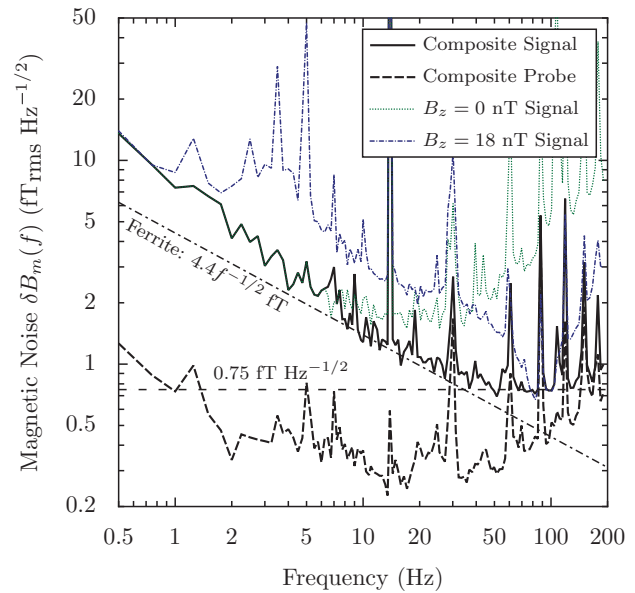}
\caption{(Color online) SERF magnetometer noise spectra. The composite noise spectrum was constructed from nine resonant spectra whose sensitivity windows covered the entire frequency range. Two such sample spectra are shown with $B_z = 0$~nT covering the low frequency range below 5 Hz and $B_z = 18$~nT covering 80 to 110 Hz range. Probe noise is obtained with unpolarized atoms. Lines indicate the calculated ferrite noise $4.4 f^{-1/2}$~fT and a baseline white noise of $0.75 \ \mathrm{fT} \ \mathrm{Hz}^{-1/2}$.}
\label{fig:bestnoise} 
\end{figure}

We measured noise spectra from the magnetometer at nine resonant frequencies in order to determine the magnetic noise spectrum up to 200 Hz. Each successive spectrum traced out a smooth envelope of noise near resonance and was dominated by probe noise off resonance. The consistency of these successive spectra indicates that the magnetometer is primarily measuring magnetic field noise. In contrast, the voltage spectrum of probe beam noise, which is due to shot noise and other non-magnetic noise, is not dependent on $B_z$. A composite spectrum shown in Fig.~\ref{fig:bestnoise} is assembled from the most responsive frequency ranges of each of the measured resonant spectra. This composite spectrum shows magnetic noise that decreases up to 35~Hz, in good agreement with the calculation of ferrite magnetization noise presented below. Deviation from the ferrite noise at very low frequencies is likely due to thermal and mechanical drift. Note that the ferrite temperature rose to 50 C during operation. Above 35~Hz, the magnetic noise levels off at $0.75 \ \mathrm{fT} \ \mathrm{Hz}^{-1/2}$ due to decreasing magnetometer signal and is primarily limited by photon shot noise.

Ferrite and $\mu$-metal, as all dissipative materials, exhibit thermal noise according to the fluctuation-dissipation theorem.\cite{Sidles:2003} Following the generalized Nyquist relation,\cite{Callen:1951} the magnetic field noise spectrum at a point inside a dissipative (non-superconducting) magnetic shield can be calculated from the power loss in the shield material generated by an oscillating current flowing in a hypothetical excitation coil located at the same point. For a ferrite shield, which has low electrical conductivity ($\sigma < 1 \ \Omega^{-1} \mathrm{m}^{-1}$), the dissipation power $P$ at low frequencies ($\ll 100$~kHz) is dominated by the hysteresis loss\cite{Beatrice:2006,Han:1995} and is given by
\begin{equation}
P=\int_V \frac{1}{2}\omega \mu ^{\prime \prime }H^{2}dV,
\label{eqn:power}
\end{equation}
where $\omega=2\pi f$ is the driving angular frequency, $\mu^{\prime\prime}$ is the imaginary part of the complex permeability $\mu=\mu^{\prime}-i\mu^{\prime\prime}$, and $H = B/|\mu|$ is the amplitude of the magnetic field intensity in the shield. The integral is carried out over the volume $V$ of the shield material.

This power loss represents an effective resistance in the excitation coil according to $P = I^2 R_\mathit{eff}/2$, where $I$ is the amplitude of the current in the coil. The effective resistance $R_\mathit{eff}$ then gives a Johnson noise voltage $\delta V=\sqrt{4 k T R_\mathit{eff}}$ across the coil. Since the only coupling between the coil and the material is through magnetic induction, this voltage noise must correspond to a magnetic field noise $\delta B_\mathit{magn}$ picked up by the coil through Faraday's law, $\delta V = A \omega \delta B_\mathit{magn}$, where $A$ is the area of the coil (taken as single-turn). As a result, the magnetic field noise due to magnetization noise in the shield is
\begin{equation}
\delta B_\mathit{magn} = \frac{ \sqrt{4kT} \sqrt{2P/I^2} }{ A\omega }
         = \sqrt{ \frac{4kT\mu ^{\prime \prime }}{\omega} } \frac{ H_\mathit{rms} \sqrt{V} }{AI},
\label{eqn:magnoise}
\end{equation}
where $H_\mathit{rms}=(\frac{1}{V}\int H^2 dV)^{1/2}$ is the rms value of $H$ in the shield material generated by a current $I$ flowing in the coil. For constant $\mu^{\prime \prime }$, the magnetization noise has a characteristic frequency dependence of $f^{-1/2}$.

The noise of our ferrite shield was determined by measuring the complex permeability of the material, MN80 from Ceramic Magnetics Inc. The permeability was determined by a four-point impedance measurement at 10--100~Hz on a sixteen-turn toroidal coil wound around the cap of the ferrite shield, a 10~mm thick circular disk with an 18~mm diameter center hole. The measurement was carried out at room temperature and with driving fields in the range of $\mu_0 H_\mathit{dr}$= 0.4--6~$\mu$T. The relative permeability was constant over the measurement frequency range and corresponded to $\mu^{\prime}/\mu_0 = 2030$ and $\mu^{\prime\prime}/\mu_0 = 6.1$ when extrapolated to the zero driving field limit. These values were taken as the low-frequency permeability of the material and used in our calculation of magnetic field noise at all frequencies ($< 200$~Hz) considered in this work. The corresponding loss factor $\mu ^{\prime\prime}\mu _{0}/\mu^{\prime 2} = 1.48 \times 10^{-6}$ was about three times smaller than the manufacturer-provided value measured at 100~kHz. The field noise at the center of the ferrite shield was then calculated from Eq.~(\ref{eqn:magnoise}) in both axial and radial directions; $H_\mathit{rms}$ was calculated numerically using commercial finite element analysis software (Maxwell 3D, Ansoft Corporation). At 1~Hz, the field noise was $\delta B_\mathit{magn} = 4.4 \pm 0.1 \ \mathrm{fT} \ \mathrm{Hz}^{-1/2}$ in both directions. The spectrum of the calculated radial noise is plotted in Fig.~\ref{fig:bestnoise}. 

The magnetic field noise due to magnetization noise in an infinitely long, cylindrical shield with $\mu \gg \mu_0$ can be analytically calculated by determining $H$ inside the shield induced by an excitation coil. Following the method for an electrostatics problem involving a charge inside a long, hollow conducting cylinder,\cite{Smythe:1968} the longitudinal magnetic field noise on the axis of the shield is found to be
\begin{equation}
\delta B_\mathit{magn} = \frac{0.26\mu _{0}}{r\sqrt{t}}\sqrt{\frac{4kT\mu ^{\prime \prime }}{\omega \mu ^{\prime 2}}},
\label{eq:hysteresisloss}
\end{equation}
where $r$ and $t$ are the inner radius and the thickness of the shield. For $r = 52$~mm, $t = 10$~mm, $T = 293$~K, this expression gives $\delta B_\mathit{magn} = 3.5 \ \mathrm{fT} \ \mathrm{Hz}^{-1/2}$ at 1~Hz.

In comparison, a $\mu$-metal shield generates magnetic field noise from both thermal magnetization noise and Johnson noise currents. Using material from a typical $\mu$-metal shield, we measured $\mu/\mu_0=3\times 10^4$ and a loss factor of $\mu ^{\prime\prime} \mu_0 / \mu^{\prime 2} = 1 \times 10^{-6}$ from impedance measurements performed at very low frequency ($< 2$ Hz). At very low frequencies, the noise is dominated by thermal magnetization noise. With Eq.~(\ref{eq:hysteresisloss}), the measurements imply thermal magnetization noise of $\delta B_\mathit{magn} = 11 f^{-1/2}$~fT for a thin shield with $t=1$~mm and $r=52$~mm, comparable in overall size to our ferrite shield. With increasing frequency, the thermal magnetization noise decreases until the Johnson noise currents dominate. These currents can be calculated from the eddy current loss;\cite{Clem:1987} for an infinitely long cylinder, analytical calculation shows that the longitudinal white noise on the axis of the cylinder is
\begin{equation}
\delta B_\mathit{eddy}=\frac{\mu _{0}\sqrt{t}}{4r}\sqrt{3kT\sigma }\cdot C(\mu),
\end{equation}
where $\sigma=1.6\times 10^{6} \ \mathrm{\Omega^{-1} m^{-1}}$ is the typical conductivity of $\mu$-metal, $C(\mu)=1$ for $\mu/\mu_0=1$, and $C(\mu)\approx 0.70$ for $\mu^\prime/\mu_0\gg 1$. The noise is white up to a roll-off frequency where self-shielding becomes dominant, due to either the skin depth effect or inductive screening. In the case of a high permeability metal, the self shielding is governed by the skin depth effect, and the noise rolls off as $f^{-1/4}$ for $f\gg 1/(\pi\mu\sigma t^2)$.\cite{Nenonen:1988} For the geometry described above, $\delta B_\mathit{eddy}=19$~fT~Hz$^{-1/2}$ at room temperature, becoming significant above 0.3~Hz and rolling off by 3~dB at about 45~Hz. We have found numerically that the transverse field noise is similar to the longitudinal field noise in our geometry, allowing valid comparison between transverse measurements and these analytical expressions.

In summary, we have demonstrated a 1~cm thick cylindrical ferrite magnetic shield with shielding properties similar to a 1~mm thick $\mu$-metal magnetic shield of the same overall size. The ferrite shield has less than half the thermal magnetization noise in the frequency range below 0.3~Hz and 25 times lower noise at 30~Hz due to the absence of Johnson noise currents. Using a ferrite shield, we have been able to demonstrate a SERF magnetometer with single-channel sensitivity of 0.75~fT Hz$^{-1/2}$, limited only by the magnetization noise of ferrite and photon shot noise. By assembling a set of resonant sensitivity measurements, we found that the primary noise in the system is magnetic and in good agreement with the calculated noise from the ferrite shield across a range of frequencies from 0 to 200~Hz. Ferrite fabrication is more economical at smaller dimensions and may be of value to compact devices that are sensitive to magnetic fields such as atomic clocks, atomic gyroscopes, magnetometers and co-magnetometers.

\vspace{12pt}

All authors contributed equally to this work. We acknowledge R. K. Ghosh for initial design of parts of the apparatus. This research was supported by DARPA.


\newpage

\begin{thebibliography}{}

\bibitem{Mager:1970}
A.~J. Mager, Magnetic Shields, IEEE Trans. Magn. \textbf{6}, 67 (1970).

\bibitem{Nenonen:1996}
J.~Nenonen,  J.~Montonen, and T.~Katila, Rev. Sci. Instrum. \textbf{67}, 2397 (1996).

\bibitem{Kominis:2003}
I.~K. Kominis, T.~W. Kornack, J.~C. Allred, and M.~V. Romalis, Nature \textbf{422}, 596 (2003).

\bibitem{Vanthull:1967}
L.~Vanthull, R.~A. Simpkins, J.~T. Harding, Phys. Lett. A \textbf{24}, 736 (1967). 

\bibitem{Vitale:1989}
S.~Vitale, G.~A. Prodi, M.~Cerdonio, J. Appl. Phys. \textbf{65}, 2130 (1989).

\bibitem{Zavattini:2006}
E.~Zavattini, G.~Zavattini, G.~Ruoso, E.~Polacco, E.~Milotti, M.~Karuza, U.~Gastaldi, G.~Di~Domenico, F.~Della Valle, R.~Cimino, S.~Carusotto, G.~Cantatore and M.~Bregant, Phys. Rev. Lett. \textbf{96}, 110406 (2006).

\bibitem{Savukov:2005}
I.~M. Savukov and M.~V. Romalis, Phys. Rev. A \textbf{71}, 023405 (2005).

\bibitem{Li:2006}
Z.~Li, R.~T. Wakai and T.~G. Walker, Appl. Phys. Lett. \textbf{89}, 134105 (2006).

\bibitem{Sidles:2003}
J.~A. Sidles, J.~L. Gabrini, W. M. Dougherty and S.-H. Chao, Proc. IEEE \textbf{91}(5), 799 (2003). 

\bibitem{Callen:1951}
H.~B. Callen and T.~A. Welton, Phys. Rev. \textbf{83}, 34 (1951).

\bibitem{Beatrice:2006}
C.~Beatrice and F.~Fiorillo, IEEE Trans. Magn. \textbf{42}, 2867 (2006).

\bibitem{Han:1995}
P.~Han, G.~R. Skutt, J.~Zhang and F.~C. Lee, in \emph{Applied Power Electronics Conference and Exposition 1995 Conference Proceedings, vol.~1}, 348 (1995).

\bibitem{Smythe:1968}
W.~R. Smythe, \emph{Static and dynamic electricity} (McGraw-Hill, New York), 188 (1968).

\bibitem{Clem:1987}
J.~Clem,  IEEE Trans. Magn. \textbf{23}, 1093 (1987).

\bibitem{Nenonen:1988}
J.~Nenonen and T.~Katila, in \emph{Biomagnetism '87}, edited by K.~Atsumi, M.~Kotani, S.~Ueno, T.~Katila and S.~J. Williamson (Denki University Press, Tokyo), 426 (1988).

\end{thebibliography}
\end{document}